\documentclass[aps,prb,twocolumn,groupedaddress]{revtex4}
\usepackage{graphics,epsfig}

\begin{document}

\title{Numerical study of resistivity of model
disordered three-dimensional metals}

\author{Yulia Gilman}
\author{Philip B. Allen}
\affiliation{Department of Physics and Astronomy, SUNY at Stony Brook, Stony Brook, NY 11794-3800}

\author{Jamil Tahir-Kheli}
\author{William A. Goddard III}
\affiliation{Material and Process Simulation Center,
Department of Chemistry,
California Institute of Technology, Pasadena, CA 91125}

\date{\today}

\begin{abstract}
We calculate the zero-temperature resistivity of model 3-dimensional  disordered metals described by tight-binding Hamiltonians. 
Two different mechanisms of disorder are considered: diagonal disorder (random on-site potentials) and off-diagonal disorder 
(random hopping integrals). The non-equilibrium Green function formalism provides a Landauer-type formula for the conductance 
of arbitrary mesoscopic systems. We use this formula to calculate the resistance of finite-size disordered samples of different 
lengths. The resistance averaged over disorder configurations is linear in sample length and resistivity is found from the
 coefficient of proportionality.  Two structures are considered: (1) a simple cubic lattice with one s-orbital per site, (2) 
a simple cubic lattice with two d-orbitals. For small values of the disorder strength, our results agree with those obtained 
from the Boltzmann equation. Large off-diagonal disorder causes the resistivity to saturate, whereas increasing diagonal disorder 
causes the resistivity to increase faster than the Boltzmann result. The crossover toward localization starts when the Boltzmann 
mean free path $\ell$ relative to the lattice constant $a$ has a value
between 0.5 and 2.0 and is strongly model dependent.
\end{abstract}

\pacs{}
\maketitle


Saturation of resistivity in metallic compounds  \cite{9} as well as its absence \cite{2} is an interesting phenomenon which 
is far from fully understood. Some compounds saturate at the levels predicted by the Ioffe-Regel condition 
($\ell = a$, where $a$ is lattice constant and $\ell$ is the mean free path), others saturate at much larger levels 
(i.e. higher resistivity), and there are some that do not saturate at all. This diversity in behavior received 
substantial attention from theorists. For example Millis {\it et al.} \cite{7} applied dynamical mean-field theory to calculate resistivity of electrons coupled
to phonons and static disorder.
Gunnarsson et al. \cite{8} studied several rather realistic models with different forms of electron-phonon coupling using quantum Monte-Carlo method.
They observed saturation of resistivity in the case of phonons coupled to hopping matrix elements.
In an attempt to understand the mechanism of saturation we chose to study transport properties of simple models of metals with static disorder (as opposed 
to models closely reproducing reality).

The resistivity was calculated using the Landauer-type formula for the zero - temperature linear response
of mesoscopic systems, which can be derived in the framework of non-equilibrium Green function formalism \cite{4}: 
$$
G = \frac{2e^2}{h} Tr[ \Gamma_L G^{\rm ret} \Gamma_R G^{\rm adv}] 
$$
where $G^{\rm ret}$ and $G^{\rm adv}$ are retarded and advanced Green's functions of the system of interest, 
 $\Gamma_L$ and $\Gamma_R$ are matrices describing the effect of contacts on the system. 

The formula above is suitable only for finite samples, whereas our aim is to calculate 
the resistivity of bulk disordered material as a function of strength of disorder.
 The solution is to calculate resistance of several samples of different lengths and 
then extract bulk resistivity $\rho$ from the data using the formula $R = L \rho /A$ ($L$ is length, $A$ is cross-section).

Let us describe the set-up of calculation for an individual sample. We consider 
a sample consisting of $N_x \times N_y \times N_z$ unit cells (for our simple
cubic examples, each unit cell 
contains one atom.) The sample is placed between two semi-infinite contacts of 
the same cross-section $N_y \times N_z$. Both contacts and sample have the same crystal
 structure and are described by a tight-binding Hamiltonian with the same parameters. 
Then tight-binding parameters for atoms inside the sample are randomly changed from their
 initial values according to rules given later and the resistance of the disordered sample is calculated. 

Our sample is not periodic in the direction of current flow $X$, but, in order to decrease the effect
 of boundaries on the results, periodic boundary conditions 
(with period $N_y$ and $N_z$) are used in the perpendicular directions
 $Y$ and $Z$. Then standard $k$-vector formalism applies in these two directions. For a given $k$-vector, Hamiltonians of the
 sample and contacts are constructed and the conductance is calculated. Then the conductance is averaged with equal weights
 over $k$-points on a uniform grid (a $6\times6$ grid was used, with a new
random Hamiltonian at each $k$-vector.)

In the limit of small disorder strength, three-dimensional transport should be accurately described 
by the linearized Boltzmann equation. Therefore, it is useful to compare our numerical results with 
resistivity obtained from the Boltzmann equation. We did not solve this equation exactly, but used 
instead the standard procedure of a displaced Fermi-Dirac distribution \cite{5} $F(k) \sim f_{FD} 
(k + eE \tau /h)$, where the displacement $eE \tau /h$ to variational accuracy is given by:
$$
\frac {h}{\tau} = 2 \pi \frac {\sum_{kk'} \overline{|V_{kk'}|^2} (v_k-v_{k'})^2 \delta (\epsilon_k -\epsilon_F) \delta (\epsilon_{k'} - \epsilon_F )} {2 \sum_k v_k^2 \delta (\epsilon_k -\epsilon_F)}
$$
where $\hbar v_k = \nabla_k \epsilon_k$, $V_{kk'}$ is the matrix element of 
scattering potential calculated in Born approximation and the bar indicates the ensemble average.

The first case is a simple cubic crystal, with one atom per unit cell and one $s$-orbital per atom. 
There are two tight-binding parameters --- the energy level $\epsilon_0$ (diagonal element of Hamiltonian) 
which is taken to be zero and the hopping integral $t=1$ between first nearest neighbors (all other hopping
 integrals are neglected). We consider a half-filled band. Two types 
of disorder are possible in this model --- diagonal and off-diagonal.

In the case of the diagonal disorder, the hopping parameter $t$ is kept constant throughout the sample but 
the energy level $\epsilon$ is changed randomly according to the formula:
$\epsilon = 0 + \xi$, where $\xi$ is a random variable distributed uniformly in $[-W/2, W/2]$. $W$ serves as 
the measure of disorder strength. The scattering potential $\overline{|V_{kk'}|^2}$ is then $W^2/12$.

In this calculation as well as in all others, the cross-section of samples is $9 \times 9$, 
and the lengths used are $5$, $6$, $7$ and $8$. For each length, $36$ configurations of disorder 
are created, the resistance of the sample for each configuration is calculated and then averaged 
over configurations. This procedure is repeated for different $L$ and $W$. In order to find the bulk
 resistivity, resistance vs. $L$ is plotted for each $W$ and resistivity is found from the slope of resulting line.

Final results are shown on Fig 1. The resistivity $\rho_B$ calculated using linearized Boltzmann equation is plotted on the same graph for comparison.
The resistivity depends linearly on $W^2$ in a quite large range of $W$ (up to $W=8$). At larger $W$ 
it deviates upwards from the Boltzmann resistivity $\rho_B$. The dotted curve shows the mean free path $\ell=<v^2>^{1/2}\tau$ 
calculated from Boltzmann resistivity  
vs $W^2$. It can be seen that deviation of resistivity from linear starts approximately when $\ell/a \sim 0.5$. This result reproduces the earlier work done by Nicolic and Allen \cite{6} (with a different computer code.)

\begin{figure}
\includegraphics[width=0.5\textwidth]{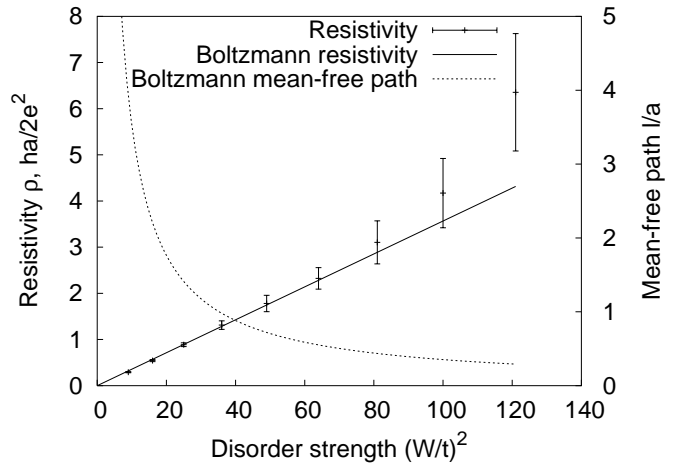}
\caption{ Resistivity for a single $s$-band model with diagonal disorder. Solid line is the Boltzmann resistivity for the same model. Dashed line shows the Boltzmann mean free path vs $W^2$.}
\end{figure}

In the case of off-diagonal disorder, the energy level $\epsilon$ is kept constant
 whereas hopping elements $t$ inside the sample are changed randomly according to: $t = 1 + \xi$, 
where $\xi$ is again a random variable distributed uniformly in $[-W/2, W/2]$. The calculational 
procedure is otherwise the same. Results for off-diagonal disorder are shown on Fig 2. It can be 
seen that resistivity depends linearly on $W^2$ up to $W=2$. At larger W, 
when $\ell/a \sim 1$, the resistivity deviates
 from linear, and at $W=4$ it starts to saturate at some level. Therefore in the case of off-diagonal disorder there is no metal-insulator transition 
(in agreement with the statement of Antoniou and Economou \cite{10}). It should also be noted that our result is similar to the results of 
Calandra and Gunnarsson \cite{3} obtained for a more realistic model of disorder induced by electron-phonon interaction. 

\begin{figure}
\includegraphics[width=0.5\textwidth]{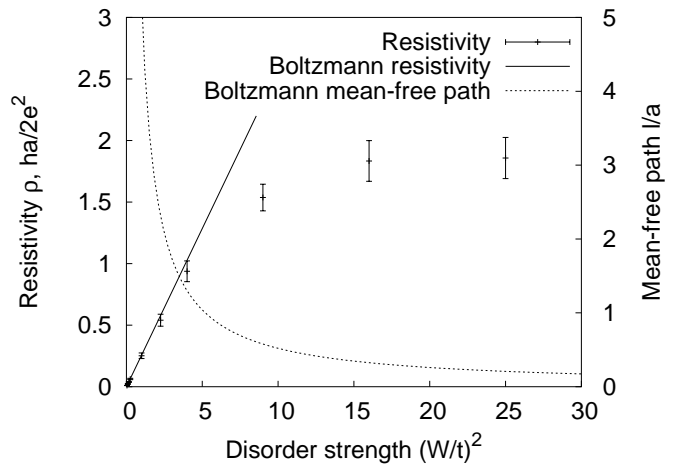}
\caption{ Resistivity for a single $s$-band model with off-diagonal disorder. Solid 
line is the Boltzmann resistivity for the same model. Dashed line shows the mean free path vs $W^2$. }
\end{figure}

We also studied the combination of diagonal and off-diagonal disorders, in order to see how interplay between them would affect resistivity.
Diagonal disorder was kept constant and off-diagonal disorder was gradually increased. Some results are shown in Fig 3. 
It is clear that the presence of diagonal disorder can change the  resistivity vs. off-diagonal disorder dependence. 
At the diagonal disorder $W_{diag}=11$ resistivity decreases with increasing off-diagonal disorder. At $W_{diag}=6$ resistivity 
displays nonmonotonic behavior. In all cases resistivity saturates eventually at approximately the same level as in the absence of static diagonal disorder (see Fig 2).

\begin{figure}
\includegraphics[width=0.5\textwidth]{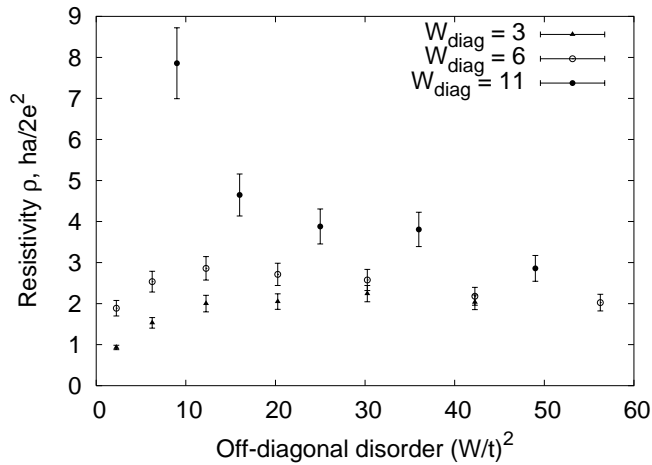}
\caption{Resistivity for a single $s$-band model with combination of disorders. Diagonal disorder is kept constant, off-diagonal disorder varies.}
\end{figure}

Next we consider a simple cubic structure with one atom per unit cell, but two $d$-orbitals per atom, 
namely $|x^2 - y^2>$ and $|3z^2  - r^2>$ ($E_g$ orbitals). As in the previous model, only first nearest
 neighbor hopping is considered. In this case, all hopping elements can be expressed in terms of two 
coupling parameters $(dd\sigma)$ and $(dd\delta)$ which are taken to be $0.051$ and $0.003 eV$ respectively. 
Both orbitals have energy levels set to zero.

First, diagonal disorder with a uniform distribution was considered. 
The two $E_g$ orbitals are given independent random diagonal energies.
Disorder strength $W$ is measured
 in terms of the full bandwidth, which for present choice of parameters equals $0.3 eV$.
The resulting graph of resistivity vs $W^2$ is shown in Fig 4 along with the Boltzmann resistivity. 
As in the case of a single orbital model, the resistivity is initially linear in $W^2$ and then begins to deviate upwards when $\ell/a\sim 1$, indicating 
the presence of a metal-insulator transition at larger $W$. 

\begin{figure}
\includegraphics[width=0.5\textwidth]{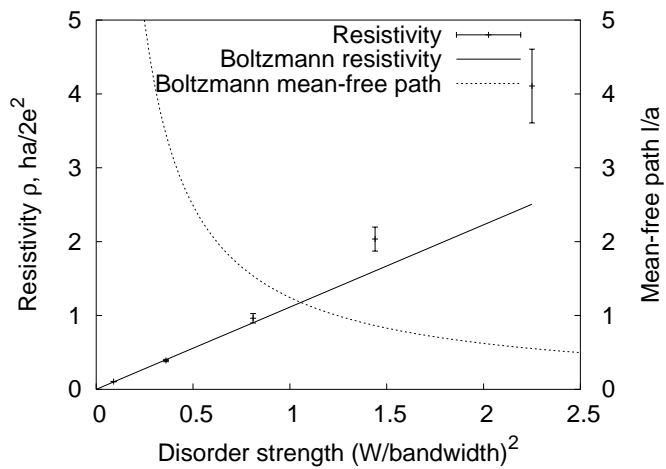}
\caption{Resistivity for the $E_g$ model with diagonal disorder. Solid line 
is the Boltzmann resistivity for the same model. Dashed line shows the mean free path 
vs $W^2$. It can be seen that resistivity deviates from Boltzmann expression when l becomes close to lattice constant a.}
\end{figure}

Off-diagonal disorder in the case of two $d$-orbital model can be created in numerous ways since
 there are several different hopping elements in the Hamiltonian. We chose disorder strength 
to be proportional to the hopping element itself, or more specifically: $t - t_0 = t_0*\xi$, where
 $\xi$ is defined above. Results are shown in Fig 5 and are similar to those of the single $s$-orbital model.
 The resistivity  starts to deviate from Boltzmann-like when
$\ell/a\sim 2$, and to saturate when $\ell$ comes close to the lattice constant, in agreement with Ioffe-Regel condition.

\begin{figure}
\includegraphics[width=0.5\textwidth]{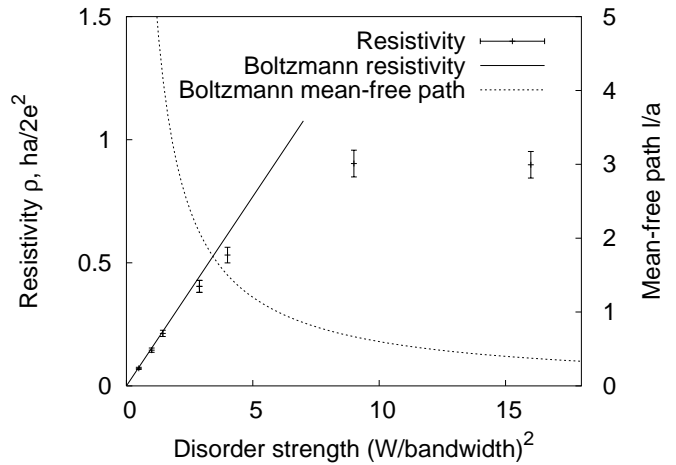}
\caption{ Resistivity for the $E_g$ model with off-diagonal disorder. Solid line is the Boltzmann resistivity for the same model. Dashed line shows the mean free path vs $W^2$.
It can be seen that resistivity deviates from Boltzmann expression when l becomes close to lattice constant a.}
\end{figure}

The main conclusion to be drawn is that the key factor to resistivity saturation in metals is strong off-diagonal disorder. 
The combination of diagonal and off-diagonal disorder can produce various types of the resistivity behaviour --- decreasing, increasing or nonmonotonic. 
Saturation is found even in the case of a single band, which is in disagreement with the statement of Allen and Chakraborty \cite{11} that multi-band 
structure is essential for saturation.  The value of $\ell/a$ at which
Boltzmann theory starts to break down varies surprisingly strongly, 
from 0.5 to 2.0, for the four models considered here.

Financial support for YG and PBA was provided by NSF (DMR-0089492).
Financial support for JTK and WAG was provided by MARCO-FENA and by NSF (DMR-0120967).



\begin{thebibliography}{99}

\bibitem{9} Z. Fisk and G.W. Webb, Phys. Rev. Lett. {\bf 36}, 1084 (1976).

\bibitem{2}  M. Gurvitch and A.T. Fiory, Phys. Rev. Lett. {\bf 59},1337 (1987); A.F. Hebard et al., Phys. Rev. B {\bf 48}, 9945 (1993).
 
\bibitem{7} A.J. Millis, Jun Hu, S. Das Sarma, Phys. Rev. Let. {\bf 82}, 2354 (1999).

\bibitem{8} M. Calandra and O. Gunnarsson, Phys. Rev. B {\bf 66} , 205105 (2002);  M. Calandra and O. Gunnarsson, Europhys. Lett. {\bf 61}, 88 (2003); O. Gunnarsson and J.E. Han, Nature {\bf 405} , 1027 (2000). 

\bibitem{4} Y. Meir and N.S. Wingreen, Phys. Rev. Let. {\bf 68}, 2512 (1992).

\bibitem{5} P.B. Allen, in {\it Quantum Theory of Real Materials}, edited by J. R. Chelikowsky and S. G. Louie
(Kluwer, Boston, 1996) Ch.17 p. 219.

\bibitem{6} B.K. Nikolic and P. B. Allen, Phys. Rev. B {\bf 63}, 020201 (2001).

\bibitem{10} P.D.Antoniou  and E.N. Economou, Phys. Rev. B {\bf 16}, 3768 (1977).

\bibitem{3} O. Gunnarsson, M. Calandra and J.E. Han, Rev. Mod. Phys. {\bf 75} , 1085 (2003). 

\bibitem{11} P.B. Allen and B. Chakraborty, Phys. Rev. Lett. {\bf 42}, 736 (1979).


\end{thebibliography}
\end{document}